\documentclass[prl,twocolumn,showpacs,superscriptaddress,floatfix]{revtex4}

\documentclass[prl,twocolumn,showpacs,superscriptaddress,floatfix]{revtex4}

\usepackage{slashed}
\usepackage{epsfig}
\usepackage{amsmath}
\usepackage{comment}
\usepackage{longtable}
\usepackage{subfigure}

\newcommand{\ppbar}{$p\bar{p}$}

\newcommand{\bbbar}{$b\bar{b}$}
\newcommand{\fb}{fb$^{-1}$}
\newcommand{\BRhbb}{$\cal{B}$($H \rightarrow b\bar{b}$)}
\newcommand{\WHlvbb}{$WH \rightarrow \ell \nu b\bar{b}$}

\newcommand{\gevcc}{GeV/$c^2$}
\newcommand{\gevc}{GeV/$c$}
\newcommand{\MET}{\mbox{$\raisebox{.3ex}{$\not\!$}E_T$}}
\newcommand{\METVEC}{\mbox{$\raisebox{.3ex}{$\not\!$}{\vec E}_T$}}

\def\ST {\texttt{SECVTX}}
\def\JP {\texttt{JETPROB}}
\def\SD {\texttt{SD}}
\def\NN {\texttt{NN}}
\def\MEBDT {\texttt{MEBDT}}

\begin{document}

\title{Search for a Higgs Boson Produced in Association with a \\ \boldmath{$W$}
 Boson in \boldmath{\ppbar} Collisions at \boldmath{$\sqrt{s} = 1.96$} TeV }

  \affiliation{Institute of Physics, Academia Sinica, Taipei, Taiwan 11529, Republic of China} 
\affiliation{Argonne National Laboratory, Argonne, Illinois 60439} 
\affiliation{University of Athens, 157 71 Athens, Greece} 
\affiliation{Institut de Fisica d'Altes Energies, Universitat Autonoma de Barcelona, E-08193, Bellaterra (Barcelona), Spain} 
\affiliation{Baylor University, Waco, Texas  76798} 
\affiliation{Istituto Nazionale di Fisica Nucleare Bologna, $^v$University of Bologna, I-40127 Bologna, Italy} 
\affiliation{Brandeis University, Waltham, Massachusetts 02254} 
\affiliation{University of California, Davis, Davis, California  95616} 
\affiliation{University of California, Los Angeles, Los Angeles, California  90024} 
\affiliation{University of California, San Diego, La Jolla, California  92093} 
\affiliation{University of California, Santa Barbara, Santa Barbara, California 93106} 
\affiliation{Instituto de Fisica de Cantabria, CSIC-University of Cantabria, 39005 Santander, Spain} 
\affiliation{Carnegie Mellon University, Pittsburgh, PA  15213} 
\affiliation{Enrico Fermi Institute, University of Chicago, Chicago, Illinois 60637}
\affiliation{Comenius University, 842 48 Bratislava, Slovakia; Institute of Experimental Physics, 040 01 Kosice, Slovakia} 
\affiliation{Joint Institute for Nuclear Research, RU-141980 Dubna, Russia} 
\affiliation{Duke University, Durham, North Carolina  27708} 
\affiliation{Fermi National Accelerator Laboratory, Batavia, Illinois 60510} 
\affiliation{University of Florida, Gainesville, Florida  32611} 
\affiliation{Laboratori Nazionali di Frascati, Istituto Nazionale di Fisica Nucleare, I-00044 Frascati, Italy} 
\affiliation{University of Geneva, CH-1211 Geneva 4, Switzerland} 
\affiliation{Glasgow University, Glasgow G12 8QQ, United Kingdom} 
\affiliation{Harvard University, Cambridge, Massachusetts 02138} 
\affiliation{Division of High Energy Physics, Department of Physics, University of Helsinki and Helsinki Institute of Physics, FIN-00014, Helsinki, Finland} 
\affiliation{University of Illinois, Urbana, Illinois 61801} 
\affiliation{The Johns Hopkins University, Baltimore, Maryland 21218} 
\affiliation{Institut f\"{u}r Experimentelle Kernphysik, Universit\"{a}t Karlsruhe, 76128 Karlsruhe, Germany} 
\affiliation{Center for High Energy Physics: Kyungpook National University, Daegu 702-701, Korea; Seoul National University, Seoul 151-742, Korea; Sungkyunkwan University, Suwon 440-746, Korea; Korea Institute of Science and Technology Information, Daejeon, 305-806, Korea; Chonnam National University, Gwangju, 500-757, Korea} 
\affiliation{Ernest Orlando Lawrence Berkeley National Laboratory, Berkeley, California 94720} 
\affiliation{University of Liverpool, Liverpool L69 7ZE, United Kingdom} 
\affiliation{University College London, London WC1E 6BT, United Kingdom} 
\affiliation{Centro de Investigaciones Energeticas Medioambientales y Tecnologicas, E-28040 Madrid, Spain} 
\affiliation{Massachusetts Institute of Technology, Cambridge, Massachusetts  02139} 
\affiliation{Institute of Particle Physics: McGill University, Montr\'{e}al, Qu\'{e}bec, Canada H3A~2T8; Simon Fraser University, Burnaby, British Columbia, Canada V5A~1S6; University of Toronto, Toronto, Ontario, Canada M5S~1A7; and TRIUMF, Vancouver, British Columbia, Canada V6T~2A3} 
\affiliation{University of Michigan, Ann Arbor, Michigan 48109} 
\affiliation{Michigan State University, East Lansing, Michigan  48824}
\affiliation{Institution for Theoretical and Experimental Physics, ITEP, Moscow 117259, Russia} 
\affiliation{University of New Mexico, Albuquerque, New Mexico 87131} 
\affiliation{Northwestern University, Evanston, Illinois  60208} 
\affiliation{The Ohio State University, Columbus, Ohio  43210} 
\affiliation{Okayama University, Okayama 700-8530, Japan} 
\affiliation{Osaka City University, Osaka 588, Japan} 
\affiliation{University of Oxford, Oxford OX1 3RH, United Kingdom} 
\affiliation{Istituto Nazionale di Fisica Nucleare, Sezione di Padova-Trento, $^w$University of Padova, I-35131 Padova, Italy} 
\affiliation{LPNHE, Universite Pierre et Marie Curie/IN2P3-CNRS, UMR7585, Paris, F-75252 France} 
\affiliation{University of Pennsylvania, Philadelphia, Pennsylvania 19104}
\affiliation{Istituto Nazionale di Fisica Nucleare Pisa, $^x$University of Pisa, $^y$University of Siena and $^z$Scuola Normale Superiore, I-56127 Pisa, Italy} 
\affiliation{University of Pittsburgh, Pittsburgh, Pennsylvania 15260} 
\affiliation{Purdue University, West Lafayette, Indiana 47907} 
\affiliation{University of Rochester, Rochester, New York 14627} 
\affiliation{The Rockefeller University, New York, New York 10021} 
\affiliation{Istituto Nazionale di Fisica Nucleare, Sezione di Roma 1, $^{aa}$Sapienza Universit\`{a} di Roma, I-00185 Roma, Italy} 

\affiliation{Rutgers University, Piscataway, New Jersey 08855} 
\affiliation{Texas A\&M University, College Station, Texas 77843} 
\affiliation{Istituto Nazionale di Fisica Nucleare Trieste/Udine, $^{bb}$University of Trieste/Udine, Italy} 
\affiliation{University of Tsukuba, Tsukuba, Ibaraki 305, Japan} 
\affiliation{Tufts University, Medford, Massachusetts 02155} 
\affiliation{Waseda University, Tokyo 169, Japan} 
\affiliation{Wayne State University, Detroit, Michigan  48201} 
\affiliation{University of Wisconsin, Madison, Wisconsin 53706} 
\affiliation{Yale University, New Haven, Connecticut 06520} 
\author{T.~Aaltonen}
\affiliation{Division of High Energy Physics, Department of Physics, University of Helsinki and Helsinki Institute of Physics, FIN-00014, Helsinki, Finland}
\author{J.~Adelman}
\affiliation{Enrico Fermi Institute, University of Chicago, Chicago, Illinois 60637}
\author{T.~Akimoto}
\affiliation{University of Tsukuba, Tsukuba, Ibaraki 305, Japan}
\author{B.~\'{A}lvarez~Gonz\'{a}lez$^r$}
\affiliation{Instituto de Fisica de Cantabria, CSIC-University of Cantabria, 39005 Santander, Spain}
\author{S.~Amerio$^w$}
\affiliation{Istituto Nazionale di Fisica Nucleare, Sezione di Padova-Trento, $^w$University of Padova, I-35131 Padova, Italy} 

\author{D.~Amidei}
\affiliation{University of Michigan, Ann Arbor, Michigan 48109}
\author{A.~Anastassov}
\affiliation{Northwestern University, Evanston, Illinois  60208}
\author{A.~Annovi}
\affiliation{Laboratori Nazionali di Frascati, Istituto Nazionale di Fisica Nucleare, I-00044 Frascati, Italy}
\author{J.~Antos}
\affiliation{Comenius University, 842 48 Bratislava, Slovakia; Institute of Experimental Physics, 040 01 Kosice, Slovakia}
\author{G.~Apollinari}
\affiliation{Fermi National Accelerator Laboratory, Batavia, Illinois 60510}
% ...

\collaboration{CDF Collaboration\footnote{With visitors from $^a$University of Massachusetts Amherst, Amherst, Massachusetts 01003,
$^b$Universiteit Antwerpen, B-2610 Antwerp, Belgium, 
$^c$University of Bristol, Bristol BS8 1TL, United Kingdom,
$^d$Chinese Academy of Sciences, Beijing 100864, China, 
$^e$Istituto Nazionale di Fisica Nucleare, Sezione di Cagliari, 09042 Monserrato (Cagliari), Italy,
$^f$University of California Irvine, Irvine, CA  92697, 
$^g$University of California Santa Cruz, Santa Cruz, CA  95064, 
$^h$Cornell University, Ithaca, NY  14853, 
$^i$University of Cyprus, Nicosia CY-1678, Cyprus, 
$^j$University College Dublin, Dublin 4, Ireland,
$^k$Royal Society of Edinburgh/Scottish Executive Support Research Fellow,
$^l$University of Edinburgh, Edinburgh EH9 3JZ, United Kingdom, 
$^m$Universidad Iberoamericana, Mexico D.F., Mexico,
$^n$Queen Mary, University of London, London, E1 4NS, England,
$^o$University of Manchester, Manchester M13 9PL, England, 
$^p$Nagasaki Institute of Applied Science, Nagasaki, Japan, 
$^q$University of Notre Dame, Notre Dame, IN 46556,
$^r$University de Oviedo, E-33007 Oviedo, Spain, 
$^s$Texas Tech University, Lubbock, TX  79409, 
$^t$IFIC(CSIC-Universitat de Valencia), 46071 Valencia, Spain,
$^u$University of Virginia, Charlottesville, VA  22904,
$^{cc}$On leave from J.~Stefan Institute, Ljubljana, Slovenia, 
}}
\noaffiliation

\date{\today}

\begin{abstract}

   We present a search for a standard model Higgs boson produced in
   association with a $W$ boson using 2.7~\fb\ of integrated
   luminosity of \ppbar\ collision data taken at $\sqrt{s} = 1.96$
   TeV.  Limits on the Higgs boson production rate are obtained for
   masses between 100 \gevcc\ and 150 \gevcc.  Through the use of
   multivariate techniques, the analysis achieves an observed
   (expected) 95\% confidence level upper limit of 5.6 (4.8) times the
   theoretically expected production cross section for a standard
   model Higgs boson with a mass of 115~\gevcc.

\end{abstract}


\maketitle

   The standard model (SM) of particle physics has proven to be
   an extremely successful theory through its accurate predictions of
   many experimental results over the last few decades.  Although the
   Higgs mechanism~\cite{Higgs:1964pj} was proposed in the 1960's, the
   fundamental particle it predicts, the Higgs boson, has yet to be
   discovered.  The mass of the Higgs boson is a free parameter of the
   SM.  However, direct limits from the LEP experiments exclude Higgs
   boson masses below $114.4$~\gevcc~\cite{Barate:2003sz} at 95\%
   confidence level (C.L.).  Taking into account additional
   electroweak precision measurements places a 95\% C.L. upper limit
   on the mass of a SM Higgs boson of
   $185$~\gevcc~\cite{Collaboration:2008ub}.  Recently, combined
   results from the CDF and D0 experiments have excluded at the 95\%
   C.L. Higgs boson masses between 160 and 170~\gevcc~\cite{Phenomena:2009pt}.

   For Higgs boson masses below $135$~\gevcc, \bbbar\ is the main
   decay mode~\cite{Djouadi:1997yw}.  In this decay, each $b$ quark
   fragments into a jet of hadrons and the Higgs boson signal may be
   reconstructed as a peak in the invariant mass distribution of these
   two jets.  At the Tevatron associated production with a $W$ boson
   ($WH$), where the $W$ boson decays into a lepton ($\ell$) and a
   neutrino ($\nu$), provides the most sensitive search channel in
   this mass range, since the requirements of a charged lepton
   candidate and of large missing transverse energy dramatically
   reduce the backgrounds from multijet processes.  Both Tevatron
   experiments, CDF and D0, have published search results for
   \WHlvbb~\cite{Aaltonen:2008zx,Abazov:2008eb,Aaltonen:2009zz}.  Here
   we describe a new search for the Higgs boson in the \WHlvbb\
   channel with increased signal acceptance that employs improved
   analysis technique and 2.7~\fb\ of \ppbar\ collision luminosity
   collected by the CDF experiment.

   The CDF II apparatus~\cite{Acosta:2004yw,geometry} is a
   general-purpose detector located at the Tevatron collider at
   Fermilab.  The detector consists of a solenoidal charged-particle
   spectrometer which includes a silicon micro-strip detector array
   surrounded by a cylindrical drift chamber in a 1.4~T axial magnetic
   field.  Outside the tracking chambers, the energies of electrons
   and jets are measured with segmented sampling calorimeters.
   Surrounding the calorimeters are layers of steel instrumented with
   planar drift chambers and scintillators used for muon
   identification.

  Events are collected with energetic lepton triggers that require one
  of the following signatures~\cite{Abulencia:2006kv}: a high-$p_T$
  electron candidate, a high-$p_T$ muon candidate, or missing
  transverse energy ($\MET$ from the neutrino escaping detection) with
  an energetic forward ($|\eta|>1.2$) electromagnetic cluster
  (designed to accept forward electrons from the $W$ boson decay).  An
  additional trigger is included that does not explicitly require an
  identified lepton, but instead requires large $\MET$ plus two
  well-separated jets in $\eta-\phi$ space~\cite{Aaltonen:2008mi}.
  For these events, the charged lepton from the $W$ boson decay is
  reconstructed only as a high-$p_T$ isolated track.  The addition of
  this non-triggered lepton category increases \WHlvbb\ signal
  acceptance by approximately 25\%~\cite{Slaunwhite:2009}.

Candidate events are selected by requiring a lepton candidate
(triggered lepton or isolated track) with
$p_T^{\ell}>20$~\gevc,~~$\MET>20$~GeV, and two jets with $|\eta|<2.0$
and $E_T>20$~GeV after correcting for instrumental
effects~\cite{Bhatti:2005ai}.  At least one of the jets must have a
displaced vertex ($b$ tag) defined by the
\ST\ algorithm~\cite{Acosta:2004hw} signaling that the jet likely
originated from a $b$ quark.  An additional $b$-tagging algorithm that
relies on high-impact-parameter tracks within jets,
\JP~\cite{Abulencia:2006kv}, is used to increase the acceptance for
double-tagged events.  Vetoes are applied to remove events with more
than one lepton and events without leptonic $W$ boson
decays~\cite{Aaltonen:2009zz}.

 The Higgs boson events are modeled with the {\sc
 pythia}~\cite{PYTHIA} MC generator combined with a parametrized
 response of the CDF II detector~\cite{GEANT3,GFLASH} and tuned to the
 Tevatron underlying event data~\cite{TUNEA}.  After basic event
 selection, the total expected signal event yield in the current
 dataset is $5.1 \pm 0.5 $~($3.5 \pm 0.4$) single (double)-tag events
 for a Higgs boson with a mass of 115~\gevcc~(see
 Table~\ref{tab:limits} for other masses).

  Models for background processes are derived from a mixture of MC
  simulation and data-driven techniques~\cite{Aaltonen:2009zz}.
  Important backgrounds to \WHlvbb\ include events with a $W$ or $Z$
  boson produced in association with jets.  These processes may
  include true $b$ jets as in $W+b\bar{b}$, or other jets that have
  been misidentified as $b$ jets like $W+c\bar{c}$ and $W+jj$ where
  $j$ refers to jets not originating from heavy-flavor quarks. Events
  with a top quark ($t{\bar{t}}$ and single top quark production),
  diboson events, and multijet events without $W$ bosons also
  contribute to the sample composition.

  After applying the event selection defined above, the background
expectation ($1896 \pm 301$ for single-tag and $316 \pm 60$ for
double-tag events) is significantly larger than the expected number of
Higgs boson signal events.  We have indicated that the dijet invariant
mass is a useful variable for separating the Higgs boson signal from
the dominant backgrounds, but the background challenge here requires
that we extract as much discrimination as possible from the full
information available in each event.  Multivariate techniques allow us to
collect the discriminating power of many variables into a single
output variable. We take advantage of the benefits from different
techniques~\cite{Aaltonen:2008ec} by combining the discriminating
power of two separate analyses that use the same event selection but
follow different multivariate strategies.  We validate the predictions
of the background model for each input variable in data control
regions.  We optimize the discriminants separately for each Higgs
boson mass hypothesis, and construct the discriminants so that they
are not sensitive to statistical fluctuations in the background and
signal samples.  We first summarize the two analyses, and then discuss
their combined result.

\begin{figure*}[th]
\begin{center}
\includegraphics[width=0.98\textwidth]{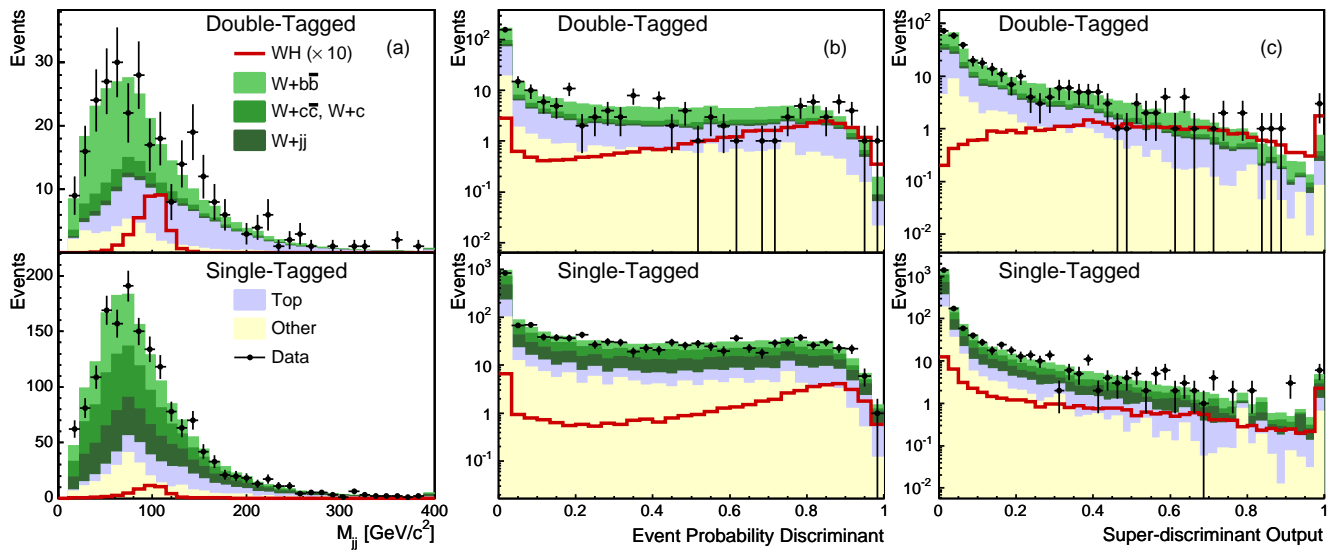}
\caption[]{The distribution for the dijet mass variable used in the \NN\
  analysis (a), the event probability discriminant used in the \MEBDT\
  analysis (b), and the \SD\ output distribution (c), for
  double $b$-tag events (top) and single $b$-tag events (bottom).  The
  background is normalized to its prediction and the signal
  expectation of a Higgs boson mass of 115~\gevcc\ is scaled to 10
  times the SM prediction.  Statistical errors are shown for the data
  points.}
\label{fig:stack}
\end{center}
\end{figure*}

The first analysis uses an artificial neural network
(\NN,~\cite{Peterson:1993nk}) trained to discriminate \WHlvbb\ signal
from the background using the information contained in the following
kinematic variables: the invariant mass of the two jets plus an
additional `loose' jet~\cite{loose} if it lies close to one of the
primary jets (angular separation $\Delta
R=\sqrt{(\Delta\eta)^2+(\Delta\phi)^2}$ less than 0.9); the vector sum
of the transverse energies,
($\sum_{jets}\vec{E}_T+\vec{p_T}^{\ell}+$\mbox{$\raisebox{.3ex}{$\not\!$}{\vec
E}_T$}); the scalar sum of the lepton and jet transverse momenta minus
the
\MET, ($\sum_{jets}E_T+{p_T}^{\ell}-\slashed E_T$); the scalar sum of
the loose jet transverse energy,
($\sum_{jets}E_T^{loose}$)~\cite{loose}; the minimum invariant mass of
the lepton, \METVEC, and one of the two jets ($\min (M_{\ell,\slashed
E_T,j_1},M_{\ell,\slashed E_T,j_2})$); and $\Delta R$ between the
lepton and the momentum of the neutrino~\cite{pZnu}.  The strongest
discriminating variable of the \NN\ is the dijet mass variable shown
in Fig.~\ref{fig:stack}(a).

The second analysis uses a boosted decision tree technique
(\MEBDT,~\cite{Hocker:2007ht,DT}).  The notation \MEBDT\ underscores
the use of inputs derived from the matrix-element approach developed
in references ~\cite{Aaltonen:2008sy,Dong:2008zzc}.  In the matrix
element method, probability densities are calculated for each event
using the measured kinematic quantities.  Some of the best
discriminating inputs to the decision tree include ratios of the
signal event probabilities to various combinations of the background
probabilities, and an event probability discriminant (EPD) defined as
the ratio of the signal event probability to the sum of the signal and
all background event probabilities as in Ref.~\cite{Aaltonen:2008sy}.
The EPD distributions for signal and backgrounds are shown in
Fig.~\ref{fig:stack}(b).

  The \MEBDT\ analysis also uses the output of a neural network that
  has been trained to separate jet flavors~\cite{Richter:2007zzc}.
  This network is based on secondary vertex tracking information and
  provides a continuous variable which helps to identify the portion
  of the background that does not contain real $b$-quark jets.  The
  \MEBDT\ analysis also includes the following inputs: the dijet mass,
  the $E_T$ of both jets and $\MET$ of the event, the difference in
  azimuthal angles ($\Delta\phi$) between the leading jet and the
  \METVEC, the $\Delta\phi$ between the lepton and the \METVEC, the
  ${p_{T}}$ and the $\eta$ of the lepton, the scalar sum of the
  transverse energies $H_T=\sum_{jets}E_T+{p_T}^{\ell}+\slashed E_T$,
  the cosine of the angle between the lepton and leading jet, and the
  transverse mass of the $W$ boson $M_{T} (W) =\sqrt{2(p_{T}^{\ell}\
  /\!\!\!\!\!E_{T}-{\vec p_{T}^{~\ell}}\!\cdot/\!\!\!\!\! {\vec
  E_{T}})}$.

  We performed the \NN\ and the \MEBDT\ analyses independently (see
  Table~\ref{tab:limits}), the results of which are partially
  correlated.  The correlations between the discriminant outputs range
  between 50\% and 75\% for the major background and signal samples.
  These correlations, while high, do suggest that a sensitivity gain
  can be obtained by combining the two approaches.  We combine the
  \NN\ and \MEBDT\ discriminants using a super-discriminant (\SD)
  technique first developed in the CDF single top quark
  search~\cite{Aaltonen:2008sy}.  Here, a neural network using the
  discriminant outputs of the \NN\ and \MEBDT\ as inputs is optimized
  using genetic
  algorithms~\cite{NEAT,Whiteson:2006ws,Aaltonen:2008bd}.  Three
  separate neural networks (one for each $b$-tag category: single \ST,
  \ST+\JP, and double \ST) are trained to separate the \WHlvbb\ signal
  from the backgrounds for each Higgs boson mass using events from the
  signal and background samples described above.  The distributions of
  the \SD\ outputs of the neural network trained for a Higgs boson
  mass of 115~\gevcc\ are shown in Fig.~\ref{fig:stack}(c) for the
  combined double-tag categories and the single-tag category.  The
  \SD\ analysis improves the sensitivity compared to the best
  individual analysis by 5-13\% for the Higgs boson masses studied.

Finding no evidence for a Higgs boson signal, we
calculate a Bayesian C.L. limit for each mass hypothesis based on the
 combined binned likelihood of the \SD\ output distributions.  The two
 lepton categories (triggered leptons and isolated tracks) and three
 tag categories yield six independent channels that are included in
 the likelihood.  Systematic uncertainties on the rate of signal and
 background production from jet energy scale, $b$-tagging
 efficiencies, lepton identification and trigger efficiencies, the
 amount of initial and final state radiation, and the parton
 distribution functions are included in the limit calculation (for
 details on systematic studies see
 ~\cite{Aaltonen:2009zz,Slaunwhite:2009}).  Uncertainties on the
 discriminant output shapes were studied but found to have a
 negligible impact on sensitivity.  A posterior density is obtained by
 multiplying this likelihood by Gaussian prior densities for the
 background normalizations and systematic uncertainties leaving
 $\sigma\times$\BRhbb\ with a uniform prior density.  A 95\%
 C.L. limit is then determined such that 95\% of the posterior density
 for $\sigma\times$\BRhbb\ falls below the
 limit~\cite{Amsler:2008zz}.  Removing systematic
 uncertainties completely from the limit calculation improves the
 expected limit by about 15\%.

\begin{figure}
\begin{center}
\includegraphics[width=\columnwidth]{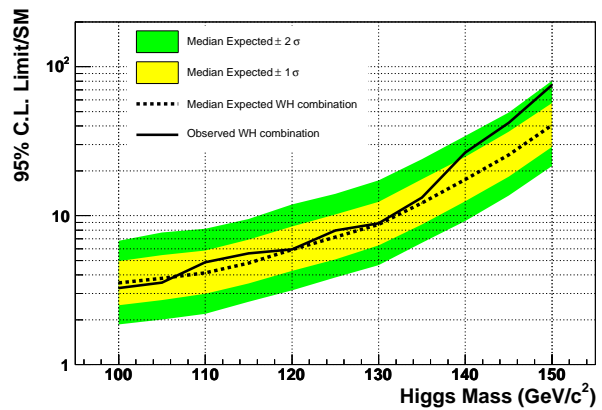}
\caption[]{The expected and observed 95\% C.L. upper
  limits on the Higgs boson production cross section relative to the
  SM expectation as obtained from the \SD\ combination as a function
  of the Higgs boson mass.}
\label{fig:limits}
\end{center}
\end{figure}

  Table~\ref{tab:limits} shows the expected and observed limits
  calculated for different Higgs boson masses.  The limits are
  displayed graphically in Fig.~\ref{fig:limits}.  We find an observed
  (expected) 95\% C.L. limit of 5.6 (4.8) times the SM prediction of
  the production cross section for a Higgs boson mass of 115~\gevcc\
  (NLO theory predicts
  $\sigma\times$\BRhbb$=136~\mbox{fb}$~\cite{Han:1991ia}).  At this
  mass, the expected limit has improved by a factor of 1.7 over the
  1.9~\fb\ result from CDF~\cite{Aaltonen:2009zz}, which corresponds
  to a 40\% improvement in sensitivity over what is expected from the
  increased dataset~\cite{lumi_note}.  The additional gain comes from
  our increased lepton acceptance through the inclusion of a
  non-triggered lepton category, a continuous jet flavor separator
  variable which improves discrimination of light-quark jets
  mistakenly tagged as $b$ jets, and the use of new multivariate
  techniques.  The excess in the observed limit at higher masses is
  due primarily to the slight excess observed at 150~\gevcc\ in the
  dijet mass variable (see Fig.~\ref{fig:stack}(a)) and is an
  indication of the large weight this variable carries in the full
  multivariate analysis.  The successful previous application of many
  of the techniques to the CDF single top
  analysis~\cite{Aaltonen:2008sy,Aaltonen:2009jj}, and the consistency
  of results obtained with \NN\ and \MEBDT\ algorithms provide further
  confidence in the robustness of the multivariate techniques. The
  increasing Tevatron dataset together with future analysis
  improvements, a combination of results from all Higgs boson
  production and decay modes, as well as the combination with the
  results from the D0 experiment~\cite{Phenomena:2009pt}, will
  continue to provide improved levels of sensitivity to the SM Higgs
  boson searches at the Tevatron.

\begin{table*}[tbh]
\begin{center}
\caption[]{The number of signal events expected to be accepted by our
selection, the SM prediction for $\sigma\times$\BRhbb, and the
expected and observed limits at 95\% C.L. on the Higgs boson
production cross section relative to the SM value as shown in
Fig.~\ref{fig:limits}.  The expected limits are also included for the
\NN\ and \MEBDT\ analyses individually.}
\begin{tabular}{l|c|c|c|c|c|c|c|c|c|c|c}\hline\hline
Mass (\gevcc)    & 100  & 105  & 110  & 115  & 120  & 125  & 130  & 135  & 140
& 145  & 150  \\ \hline
Exp. signal (events) &  12.8 & 11.7 & 10.3 & 8.6 & 6.9 & 5.6 & 4.3 & 3.1 & 2.1 & 1.4 & 0.9 \\ \hline
SM $\sigma\times$\BRhbb\ (fb) & 232 & 201 & 169 & 136 & 104 & 83 & 63 & 45 & 30 & 20 & 12 \\ \hline
Exp. NN (95\% C.L./SM)   & 4.3  & 4.6  & 5.0  & 5.8  & 6.9  & 8.2  & 10.0 & 13.8 & 19.4 & 28.9 & 43.2 \\ 
Exp. MEBDT (95\% C.L./SM) & 3.8  & 4.0  & 4.5  & 5.2  & 6.3  & 8.0  & 10.0 & 13.4 &
19.2 & 27.0 & 48.7 \\ \hline
Exp. combination (95\% C.L./SM) & 3.5  & 3.8  & 4.1  & 4.8  & 5.9  & 7.2  & 8.7  & 12.2 & 17.5 & 25.6 & 40.5 \\
Observed  (95\% C.L./SM)     & 3.3  & 3.6  & 4.9  & 5.6  & 5.9  & 8.0  & 8.9  & 13.2 & 26.5 & 42.1 & 75.5 \\
\hline\hline
\end{tabular}
\label{tab:limits}
\end{center}
\end{table*}

We thank the Fermilab staff and the technical staffs of the
participating institutions for their vital contributions. This work
was supported by the U.S. Department of Energy and National Science
Foundation; the Italian Istituto Nazionale di Fisica Nucleare; the
Ministry of Education, Culture, Sports, Science and Technology of
Japan; the Natural Sciences and Engineering Research Council of
Canada; the National Science Council of the Republic of China; the
Swiss National Science Foundation; the A.P. Sloan Foundation; the
Bundesministerium f\"ur Bildung und Forschung, Germany; the Korean
Science and Engineering Foundation and the Korean Research Foundation;
the Science and Technology Facilities Council and the Royal Society,
UK; the Institut National de Physique Nucleaire et Physique des
Particules/CNRS; the Russian Foundation for Basic Research; the
Ministerio de Ciencia e Innovaci\'{o}n, and Programa
Consolider-Ingenio 2010, Spain; the Slovak R\&D Agency; and the
Academy of Finland.   



\begin{thebibliography}{35}
\bibitem[{geo()}]{geometry}
\bibinfo{note}{We use a cylindrical coordinate system with the origin at the
  center of the CDF detector, $z$ pointing in the direction of the proton beam,
  $\theta$ and $\phi$ representing the polar and azimuthal angles,
  respectively, and pseudorapidity defined by $\eta= - \rm ln$ $\rm
  tan(\theta/2)$. The transverse momentum $p_T$ (transverse energy $E_{T}$) is
  defined to be $p\sin\theta$ ($E\sin\theta$). The missing $E_T$ ($\METVEC$) is
  defined by $\METVEC =-\sum_iE_T^i{\hat n}_i$, ${\hat n}_i$ is a unit vector
  perpendicular to the beam axis and pointing at the $i$th calorimeter tower
  ($\MET=|\METVEC|$).}
\bibitem[{loo()}]{loose}
\bibinfo{note}{'Loose' jets are defined to be exclusive to our primary jet
  definition ($|\eta|<2.0$ and $E_T>20$~GeV) with $|\eta|<2.4$ and
  $E_T>12$~GeV.}
\bibitem[{pZn()}]{pZnu}
\bibinfo{note}{The neutrino $p_z$ is chosen as the solution of the $W$ mass
  constraint equation that gives the largest $|p_z|$ for the neutrino.}
\bibitem[{lum()}]{lumi_note}
\bibinfo{note}{We assume that the sensitivity would scale inversely to the
  square root of the integrated luminosity in the absence of analysis
  improvements.}
\end{thebibliography}



\end{document}